\documentstyle[aps,tighten,preprint,floats,12pt,epsf]{revtex}
\newcommand{\logqmms}{l_{qm}}

\newcommand{\logqmums}{l_{q\mu}}

\newcommand{\gsim}{\;\rlap{\lower 3.5 pt \hbox{$\mathchar \sim$}} \raise 1pt
 \hbox {$>$}\;}
\newcommand{\lsim}{\;\rlap{\lower 3.5 pt \hbox{$\mathchar \sim$}} \raise 1pt
 \hbox {$<$}\;}

\renewcommand{\thefootnote}{\fnsymbol{footnote}}
\begin{document}    

\preprint{
\hbox to \hsize{
\hfill$\vtop{   \hbox{MPI/PhT/98--06}
                \hbox{TTP98--04}
                \hbox{hep-ph/9801432}
                \hbox{January 1998}
                \hbox{}
                \hbox{}}$}
}
\title{Singlet Polarization Functions at \boldmath{${\cal O}(\alpha_s^2)$}}
\author{K.G.~Chetyrkin\thanks{Permanent address: Institute for Nuclear
   Research, Russian Academy of Sciences, Moscow 117312, Russia.}
 and
 R.~Harlander}
\address{Institut f\"ur Theoretische Teilchenphysik,
    Universit\"at Karlsruhe,\\ D-76128 Karlsruhe, Germany}
\author{M.~Steinhauser}
\address{Max-Planck-Institut f\"ur Physik,
    Werner-Heisenberg-Institut,\\ D-80805 Munich, Germany}
\maketitle

\begin{abstract}
\noindent 
We consider the three-loop singlet diagrams induced by axial-vector,
scalar and pseudo-scalar currents.  Expansions for small and large
external momentum $q$ are presented.  They are used in combination
with conformal mapping and Pad\'e approximations in order to arrive at
results for the polarization functions valid for all $q^2$. Results are
presented for the imaginary parts which are directly related to physical
quantities like the production of top quarks or the decay of scalar or
pseudo-scalar Higgs bosons.

\end{abstract}



\renewcommand{\thefootnote}{\arabic{footnote}}
\setcounter{footnote}{0}

\section{Introduction}\label{introsect}

In the last few years a lot of effort has been devoted to
developments of techniques which allow the evaluation
of higher order corrections.
Of special interest is thereby the evaluation of QCD corrections to
two-point current correlators. Their knowledge immediately
leads to a variety of important observables like the cross section
$\sigma(e^+e^-\to\mbox{hadrons})$ mediated by a photon
or a $Z$ boson or the decay of a scalar or pseudo-scalar Higgs
boson.
Whereas at the one- and two-loop level exact results are known
(for a review see \cite{CKKRep}),
until recently at ${\cal O}(\alpha_s^2)$ only expansions
for large external momentum $q$ respectively small quark mass $m$
were available.

In \cite{BaiBro95,CheKueStePade} an approach was developed which leads
to semi-analytical results for the three-loop polarization functions.
The essence of this procedure amounts to the combination of the low- and
high-energy analytical data for a polarization function through the use
of the conformal mapping and Pad\'e approximation suggested in
\cite{FleTar94,BroFleTar93,BroBaiIly94}.

In a first step it was applied to the non-singlet diagrams induced by
external vector, axial-vector, scalar and pseudo-scalar
currents~\cite{CheKueStePade}.  In this paper this procedure will be
applied to the corresponding singlet diagrams. They are often also
referred to as double-triangle diagrams as the external currents are not
connected through the same fermion line.  This completes the knowledge
of the three-loop current correlators at ${\cal O}(\alpha_s^2)$.  Thus
also the full mass dependence for the inclusive cross sections
$\sigma(e^+e^-\to t\bar t)$ and the decays of a scalar or pseudo-scalar
Higgs boson into quarks is available at this order.

The method used in \cite{CheKueStePade} heavily relies on the fact that
the lowest particle threshold for the non-singlet graphs 
starts at $q^2=4m^2$. In contrast to that, the
singlet diagrams contain massless cuts. The solution of this problem is
described in detail for the axial-vector correlator
in Section~\ref{secax}. 
Also the essential ingredients are listed and the 
approximation procedure is briefly reviewed. 
Section~\ref{secscps} describes the treatment of
the scalar and pseudo-scalar diagrams and finally results are presented
in Section~\ref{secres}.
Note that for the vector correlator there are no singlet diagrams 
at three-loop level according to Furry's theorem.


\section{Singlet axial-vector correlator}
\label{secax}

In this section the ingredients and the procedure for the 
construction of the Pad\'e approximants
of the axial-vector polarization function are presented in detail.
The scalar and pseudo-scalar singlet diagrams are discussed
in Section~\ref{secscps}.

Let us start with some definitions. It is convenient to introduce
the dimensionless quantities
\begin{eqnarray}\label{eqdefinitions}
z\,\,=\,\,\frac{q^2}{4m^2},
\quad
r\,\,=\,\,\frac{s}{4m^2},
\quad
x\,\,=\,\,\frac{2m}{\sqrt{s}},
\quad
v\,\,=\,\,\sqrt{1-x^2},
\end{eqnarray}
where $\sqrt{s}$ is the center of mass energy and $m$ is the pole mass
of the produced quark.
$x$ is a convenient variable in the high energy region
and $v$ represents the velocity of the quark.
The axial-vector polarization function is defined through
\begin{eqnarray}
\left(-q^2g_{\mu\nu}+q_\mu q_\nu\right)\,\Pi^a(q^2)
+q_\mu q_\nu\,\Pi^a_L(q^2)
&=&
i\int dx\,e^{iqx}\langle 0|Tj^a_\mu(x) j^a_\nu(0)|0 \rangle,
\label{eqpivadef}
\end{eqnarray}
with
$j_\mu^a = \bar{\psi}\gamma_\mu\gamma_5 \psi$. Only the transversal part
$\Pi^a(q^2)$ will be considered in the following.
In $D=4-2\varepsilon$ space-time dimensions care has to be 
taken concerning the treatment of $\gamma_5$ --- especially
in connection with the singlet diagrams.
We follow the treatment introduced in
\cite{tHoVel72BreMai77} and refer for more details to
\cite{HarSte972}.

\begin{figure}[t]
 \begin{center}
 \begin{tabular}{cc}
   \leavevmode
   \epsfxsize=6.5cm
   \epsffile[131 314 481 478]{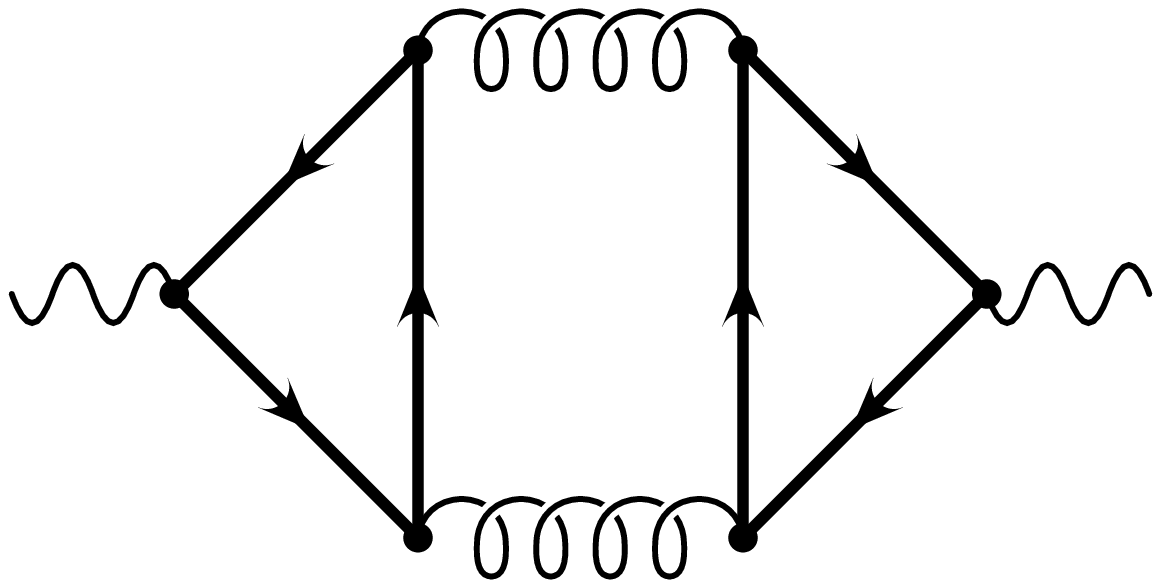}
   &
   \epsfxsize=6.5cm
   \epsffile[131 314 481 478]{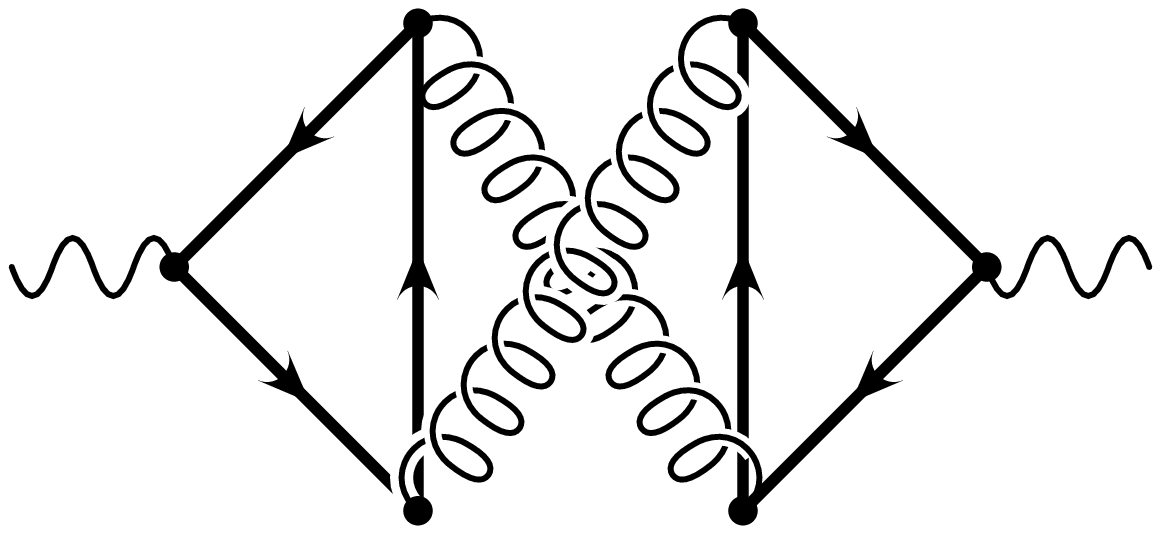}
 \end{tabular}
 \caption{\label{figdiasing}Singlet or double-triangle diagrams.
   In the fermion lines either the quark $\psi$ or its isospin partner
   $\chi$ may be present.}
 \end{center}
\end{figure}

In order for the axial anomaly to cancel, one has to take both members
of a weak isospin doublet into account. It is therefore convenient to
replace the current $j^a_\mu$ in Eq.~(\ref{eqpivadef}) by 
$
j_{S,\mu}^a = \bar{\psi}\gamma_\mu\gamma_5 \psi - 
\bar{\chi}\gamma_\mu\gamma_5 \chi ,
$
where $\psi$ and $\chi$ are isospin partners.
The diagrams contributing to the singlet part, $\Pi^a_S(q^2)$, are
depicted in Fig.~\ref{figdiasing}
where in the fermion triangles either $\psi$ or $\chi$ may be present.
Note that for a degenerate quark doublet $\Pi_S^a(q^2)$ vanishes.
Having in mind the physical case $(\psi,\chi) = (t,b)$, however, we set
$m_\psi = m$ and $m_{\chi} = 0$ in the subsequent analysis.
%
$\Pi^a_S(q^2)$ is conveniently written in the form
\begin{eqnarray}
\Pi_S^a(q^2) = C_F T \left(\frac{\alpha_s}{\pi}\right)^2
\Pi_S^{(2),a}(q^2),
\label{eqpis2}
\end{eqnarray}
where $C_F = (N_c^2-1)/(2N_c)$, and $T$ is the trace normalization for an
$SU(N_c)$ gauge group. For QCD $N_c=3$ and $T=1/2$.

The imaginary part of $\Pi_S^a(q^2)$,
\begin{eqnarray}
R_S^a (s)&=&12\pi\,\mbox{Im}\,\Pi_S^a(q^2=s+i\epsilon),
\end{eqnarray}
normalized in analogy to the vector case, enters, e.g., the total
inclusive cross section for the production of top quarks. However, also
imaginary parts arising from the massless quark $\chi$ contribute to
$R_S^a(s)$. The purely gluonic cut is zero according to the
Landau-Yang-Theorem \cite{LanYan}.
Let $R_{Sb}^{(2),a}$ be the contribution of these massless cuts to
$R_S^a$. Then the quantity
\begin{equation}\label{eqpiahat}
\hat\Pi^{a}_S(q^2) = {1\over 12\pi^2}\int_0^1 dr {R_{Sb}^{(2),a}(s)\over
  r-z}
\end{equation}
defines a function whose imaginary part for $q^2\le 4m^2$ coincides with
that of $\Pi_S^{(2),a}(q^2)$ and is zero for $q^2>4m^2$. Its evaluation
will be described below.
With the help of this function we choose the overall renormalization
condition for $\Pi_S^{(2),a}$
\begin{equation}\label{renormbeda}
\lim_{q^2\to 0}\left(\Pi_S^{(2),a}(q^2) - \hat\Pi_S^{(2),a}(q^2)\right)
= 0\,.
\end{equation}
In this difference the $\ln(-z)$ terms cancel for $z\to 0$
which makes it possible to demand this QED-like renormalization
condition. Alternatively one could use the $\overline{\mbox{MS}}$ scheme where
only the $1/\varepsilon$ poles are subtracted.

An important ingredient for the Pad\'e approximation is the expansion of 
$\Pi_S^{(2),a}(q^2)$ for small external momentum. It is obtained by
applying the so-called hard mass procedure~\cite{hmp}
which provides a consistent expansion in $q^2/m^2$:
\begin{eqnarray}
\Pi_S^{(2),a}(q^2) &=& \frac{3}{16\pi^2}\sum_{n\ge 0} C_{S,n}^{(2),a} z^n.
\label{eqpissma}
\end{eqnarray}
The first eight coefficients have been evaluated:
\begin{eqnarray}
C_{S,0}^{(2),a} &=& K^a + {37\over 6}\logqmms - \logqmms^2 \nonumber\\
C_{S,1}^{(2),a} &=&
             {471631\over 311040} 
           + {5215\over 13824}\,\zeta_3
           - {56\over 81}\,\logqmms \nonumber \\ 
C_{S,2}^{(2),a} &=&
             {1178417\over 8064000} 
           + {18179\over 92160}\,\zeta_3
           - {14\over 75}\,\logqmms \nonumber \\ 
C_{S,3}^{(2),a} &=&
           - {15527720419\over 5120962560000} 
           + {1811719\over 14745600}\,\zeta_3
           - {13312\over 165375}\,\logqmms \nonumber \\ 
C_{S,4}^{(2),a} &=&
           - {23054974995287\over 811160469504000} 
           + {5965963\over 70778880}\,\zeta_3
           - {7696\over 178605}\,\logqmms \nonumber \\ 
C_{S,5}^{(2),a} &=&
           - {735623850793897673\over 23817834479222784000} 
           + {16260413\over 264241152}\,\zeta_3
           - {6600448\over 252130725}\,\logqmms \nonumber \\ 
C_{S,6}^{(2),a} &=&
           - {1301167265336208772211\over 46002446022727434240000} 
           + {425212357\over 9059696640}\,\zeta_3
           - {3007168\over 173918745}\,\logqmms \nonumber \\ 
C_{S,7}^{(2),a} &=&
           - {11171249029492611725205473\over 450455951454547036078080000} 
           + {193004110999\over 5218385264640}\,\zeta_3
           - {132788224\over 10956880935}\,\logqmms\,, \nonumber \\
           \label{coefaq20}
\end{eqnarray}
with $\zeta_3\approx1.202056903$ and $\logqmms = \ln(-q^2/m^2)$.
$K^a$ is a constant whose numerical value will be given below.

A stringent constraint both for the real and imaginary part of
$\Pi^{(2),a}_S(q^2)$ is set by the expansion for large external
momentum. Recently, the large momentum procedure has been applied
leading to the result
\cite{HarSte972}\footnote{In \cite{HarSte972} the results are listed in the
$\overline{\mbox{MS}}$ scheme.}:
\begin{eqnarray}
  \Pi^{(2),a}_S(q^2) &=& {3\over 16\pi^2} \bigg\{
  K^a
+ \frac{185}{16} 
- \frac{385}{288}\zeta_3 
+ 21 \zeta_3 \,{m^2\over q^2}
       \nonumber\\&&\mbox{}
+    \left({m^2\over q^2}\right)^{2}\,\bigg[
           - {80\over 3}\,\zeta_3 
           + {320\over 3}\,\zeta_5
          \bigg]
       \nonumber\\&&\mbox{}
    + \left({m^2\over q^2}\right)^{3}\,\bigg[
             {380\over 3} 
           - 64\,\zeta_3
           + \left(
                 {296\over 3} 
               - 32\,\zeta_3
              \right)\,\logqmms 
           + 24\,\logqmms^2 
          \bigg] 
       \nonumber\\&&\mbox{}
    + \left({m^2\over q^2}\right)^{4}\,\bigg[
           - {3271\over 243} 
           - {416\over 9}\,\zeta_3 
           + \left(
                 {280\over 27} 
               + 32\,\zeta_3
              \right)\,\logqmms
           + {410\over 27}\,\logqmms^2 
           - {176\over 27}\,\logqmms^3 
          \bigg] 
       \nonumber\\&&\mbox{}
    + \left({m^2\over q^2}\right)^{5}\,\bigg[
           - {395921\over 2916} 
           - {5584\over 27}\,\zeta_3 
       \nonumber\\&&\mbox{\hspace{.5cm}}
           + \left(
                 {4111\over 54} 
               + {160\over 3}\,\zeta_3
              \right)\,\logqmms
           + {1340\over 9}\,\logqmms^2 
           - {1660\over 81}\,\logqmms^3 
          \bigg] 
       \nonumber\\&&\mbox{}
    + \left({m^2\over q^2}\right)^{6}\,\bigg[
           - {105441373\over 101250} 
           - {2420\over 3}\,\zeta_3 
       \nonumber\\&&\mbox{\hspace{.5cm}}
           + \left(
               - {6044237\over 40500} 
               + 112\,\zeta_3
              \right)\,\logqmms
           + {1177331\over 1350}\,\logqmms^2 
           - {15542\over 135}\,\logqmms^3 
          \bigg]
\bigg\}
+ \ldots\,\,,
\label{eqpi2as}
\end{eqnarray}
with $\zeta_3$ defined above and $\zeta_5\approx1.036927755$.

The logarithms in Eq.~(\ref{coefaq20}) are due to the massless cuts 
and do not appear in the non-singlet diagrams \cite{CheKueStePade}.
They spoil the procedure for constructing
the Pad\'e approximants developed in \cite{CheKueStePade}.
Therefore, instead of dealing with the full polarization function, 
one may use Eq.~(\ref{eqpiahat}) and consider the following quantity:
\begin{eqnarray}
  \Pi_{S,mod}^{(2),a}(q^2) &=& \Pi_S^{(2),a}(q^2) -
  \hat\Pi_S^{(2),a}(q^2)\,.
\label{eqpiamod}
\end{eqnarray}
According to the definition of $\hat\Pi_S^{(2),a}(q^2)$ the
$\ln(-z)$-terms from the low energy expansion of $\Pi_S^{(2),a}(q^2)$
are exactly canceled. Above $z=1$, the imaginary part on the r.h.s.~of
Eq.~(\ref{eqpiamod}) is determined by $\Pi_S^{(2),a}(q^2)$ alone.  This
means that for $z>1$, $\Pi_{S,mod}^{(2),a}(q^2)$ contains all possible
cuts of the double-triangle diagrams and one should subtract the
massless ones by using $R_{Sb}^{(2),a}(s)$ in the region $s>4m^2$ to get
the production cross section for massive quarks.

An analytic formula for $R_{Sb}^{(2),a}(s)$ is available
\cite{KniKue89}\footnote{\label{footrsb}In \cite{KniKue89} a different
  kinematical region was considered. However, we continued the result to
  the region under consideration by using the translation table given in
  the appendix of \cite{KniKue89}.}. Nevertheless it is not possible to
solve the dispersion integral in Eq.~(\ref{eqpiamod}) analytically.  On
the other hand a purely numerical integration is excluded as the result
contains $\ln(-z)$ terms for $z\to0$ which makes an expansion of the
integrand with subsequent integration impossible.  Let us therefore
briefly describe the method we used for evaluation of
Eq.~(\ref{eqpiahat}).

One may write $R_{Sb}^{(2),a}(s)$ as (recall Eq.~(\ref{eqdefinitions}))
\begin{equation}
R_{Sb}^{(2),a}(s) = 
3\,\frac{3}{2}\ln(4r) + \tilde{R}_{Sb}^{(2),a}(s).
\label{eqrsb}
\end{equation}
Then $\tilde{R}_{Sb}^{(2),a}(s)$ has a very simple limiting behaviour. For
$s\to 0$ it reads
\begin{equation}
\tilde{R}_{Sb}^{(2),a}(s)= 
3\left[
- \frac{37}{8} + \frac{14}{27}r + \frac{7}{50}r^2 +
\ldots
\right],
\label{eqrsbsto0}
\end{equation}
where the dots represent higher orders in $r$.
The expansion of $\tilde R_{Sb}^{(2),a}(s)$ around $\sqrt{s}=2m$ leads to
\begin{eqnarray}
  \tilde R_{Sb}^{(2),a}(s) &=& 3\left[
  -\frac{19}{8}
  -\frac{3}{2}\zeta_2
  +\frac{7}{8}\zeta_3
  +(1-r)\left(\frac{1}{2}-3\zeta_2+\frac{7}{4}\zeta_3\right)
\right.
\nonumber\\&&\mbox{}
\left.
  +\left(\sqrt{1-r}\right)^3\left(2\pi-\frac{4}{3}\pi\ln2\right)
+ \ldots
\right],
\label{eqrsbsto1}
\end{eqnarray}
i.e., in this limit $\tilde R_{Sb}^{(2),a}(s)$ is a series in 
$\sqrt{1-r}$.
Provided with this information we split the integral in
Eq.~(\ref{eqpiahat}) into three parts:
\begin{eqnarray}
\int_0^1\,dr\, \frac{R_{Sb}^{(2),a}(s)}{r-z}
&=&
\int_0^1\,dr\, \frac{3\,\frac{3}{2}\ln(4r)}{r-z}
+
\int_0^\delta\,dr\, \frac{\tilde{R}_{Sb}^{(2),a}(s)}{r-z}
+
\int_\delta^1\,dr\, \frac{\tilde{R}_{Sb}^{(2),a}(s)}{r-z}
\label{eqsplit}
\end{eqnarray}
and replace in a second step $\tilde{R}_{Sb}^{(2),a}(s)$
in the interval $[0,\delta]$ by the expansion in (\ref{eqrsbsto0}),
in the interval $[\delta,1]$ by the one in (\ref{eqrsbsto1}).
It turns out that the inclusion of the first 100 terms in the
small energy expansion and the first 40 terms in the expansion
around $r=1$ leads to stable results in the range 
$\delta=0.65 \ldots 0.80$ with an accuracy of $13$ to $14$
digits.
We will not quote numbers for the full $\hat\Pi_S^{(2),a}(q^2)$ but only
for the constant $K^a$ appearing in Eqs.~(\ref{coefaq20}) and (\ref{eqpi2as}):
\begin{equation}
K^a = -9.08040684374401\ldots\,\,.
\end{equation}

The third kinematic region to be used for the Pad\'e procedure is the
threshold for the production of two massive quarks, $z\to1$.  In this
region $\Pi_S^{(2),a}(q^2)$ gets contributions from two sources: the
cuts involving massive quarks and $R_{Sb}^{(2),a}(s)$.  It is strongly
expected that the former starts at least with a term proportional to $v$
in analogy to the non-singlet axial-vector correlator which follows the
$P$-wave scattering solution of the Coulomb potential.
$R_{Sb}^{(2),a}(s)$ on the other hand has a smooth behaviour for
$s\to4m^2$.  For $z<1$, $\Pi_{S,mod}^{(2),a}(q^2)$ is constructed in such
a way that its imaginary part vanishes. However, the leading
contribution of $\Pi_{S,mod}^{(2),a}(q^2)$ for $z\to1^+$ is given by
\begin{equation}
  \Pi_{S,mod}^{(2),a}(q^2) =
  \frac{3}{16\pi^2}\ln\left(\frac{1}{1-z}\right)
  \left(-\frac{19}{6}+4\ln2-2\zeta_2+\frac{7}{6}\zeta_3\right)
    +\ldots\,\,,
\label{eqpismodthr}
\end{equation}
where the ellipses represent sub-leading terms in $(1-z)$.

The construction of the Pad\'e approximations divides naturally into
four steps (for more detail we refer to \cite{CheKueStePade}): First,
the threshold contribution has to be subtracted in all kinematical
regions in order to have a polarization function that has a vanishing
imaginary part for $z\to1$.  Then, a new polarization function,
$\tilde{\Pi}_{S,mod}^{(2),a}(q^2)$, is constructed whose high
energy expansion contains no logarithmic terms any more. This must be
done carefully in order not to destroy the behavior for $z\to0$
and $z\to1$.  In a third step the conformal mapping~\cite{FleTar94}
\begin{equation}\label{conformmap}
z = {4\omega\over (1+\omega)^2}
\end{equation}
is used to transform the $q^2$ plane into the interior of the unit
circle.  Finally, a Pad\'e improvement is performed in the new variable,
$\omega$.

In \cite{CheKueStePade} only the constant and the $m^2/q^2$ corrections
in the high energy expansion have been included into the analysis.
For the singlet diagrams meanwhile terms up to 
${\cal O}((m^2/q^2)^{6})$ are available. This makes it necessary to modify
the definition of the function $P(\omega)$ for which
the Pad\'e approximation is performed.
The natural extension of the definition given in
\cite{CheKueStePade} reads:
\begin{eqnarray}
P_n(\omega) &=&  {(4\omega)^{n-1}\over (1+\omega)^{2n}}\left(
  \tilde\Pi_{S,mod}^{(2),a}(q^2) - 
  \sum_{j=0}^{n-1}{1\over j!}\left(
  {d^j\over d(1/z)^j}\tilde\Pi_{S,mod}^{(2),a}(q^2)\bigg|_{z =
  -\infty}\right) {(1+\omega)^{2j}\over (4\omega)^j}\right)\,,
\end{eqnarray}
where the index $n\geq 1$ indicates that the mass
corrections of order $(m^2/q^2)^n$ are included.  
The Pad\'e approximants
\begin{equation}
[i/j](\omega) = {a_0 + a_1\omega +\cdots + a_i\omega^i\over
  1 + b_1\omega + \cdots + b_j\omega^j}
\end{equation}
are then constructed from $P_n(-1)$ and $P_n^{(k)}(0), (k =
0,1,\ldots,n+n_0-1)$, where $n_0$ is the number of moments (see
Eq.~(\ref{coefaq20})) used for the construction of the Pad\'e
approximation and $P_n^{(k)}(0) = {d^k\over
  d\omega^k}P_n(\omega)|_{\omega=0}$.  Taking into account all available
information, i.e.~$n=6$ and $n_0=7$, it is possible to construct
approximants like $[7/6]$, $[6/7]$ or $[8/5]$.  However, it turns out
that the construction of $P_n^{(k)}(0)$ for large values of $k$ suffers
from huge cancellations. It is therefore necessary to evaluate the
expressions in Eq.~(\ref{eqsplit}) with highest possible accuracy in
order to arrive at reliable results for high-order Pad\'e
approximations.  After all, the above mentioned $13$ to $14$ digits are
enough to get stable results. Of course, also lower order Pad\'e's have
been evaluated both for consistency checks and to examine the
convergence properties.  We should mention that some of the Pad\'e
approximants develop poles for $|\omega|<1$ which result in poles in
the physical $z$ plane. Since this is not acceptable, only Pad\'e
approximants free from poles in the physical region are considered in
the discussion of Section \ref{secres}.  We refrain from listing
explicit formulae for $\Pi_S^{(2),a}(q^2)$ at this point and instead
present in Section~\ref{secres} results for the imaginary part,
$R^{(2),a}_S(s)$.


\section{Scalar and pseudo-scalar case}
\label{secscps}

It is now straightforward to extend the procedure described above to the
scalar and pseudo-scalar case.  Here, in contrast to the singlet
axial-vector contribution only the diagram with two massive triangles
contributes. However, the cut through the two gluons does not vanish as
it was the case for the axial-vector coupling, so that again there is a
cut starting at $z=0$.

The polarization functions are defined through 
($\kappa$ stands for $s$ and $p$, denoting the scalar and pseudo-scalar
case, respectively):
\begin{eqnarray}
q^2\,\Pi^\kappa(q^2)
&=&
i\int dx\,e^{iqx}
\langle 0|Tj^\kappa(x)j^\kappa(0)|0 \rangle,
\label{eqpispdef}
\\
R^\kappa (s)&=&8\pi\,\mbox{Im}\,\Pi^\kappa(q^2=s+i\epsilon),
\label{eqrtopisp}
\end{eqnarray}
where the currents are given by $j^s = \bar{\psi}\psi$ and $j^p = i
\bar{\psi}\gamma_5 \psi$. As for $j^p$ one again has to deal with
$\gamma_5$ in $D \neq 4$ dimensions, we again adopt the definition of
\cite{tHoVel72BreMai77}, referring for details to~\cite{HarSte97}.

In analogy to (\ref{eqpiahat}) we define:
\begin{equation}\label{eqpikhat}
  \hat\Pi^{\kappa}_S(q^2) = {1\over 8\pi^2}\int_0^1 dr
  {R_{gg}^{(2),\kappa}(s)\over r-z}\,.
\end{equation}
$R_{gg}^{(2),\kappa}(s)$ corresponds to the two gluon cut
actually describing the Born decay of a scalar or pseudo-scalar
Higgs boson to gluons \cite{EllGaiNan76,DjoSpiZer93}:
\begin{eqnarray}
R_{gg}^{(2),s} \,\,=\,\, 
\frac{3}{2r}\left(1+\frac{r-1}{r}f(r)\right)^2,
&\qquad&
R_{gg}^{(2),p} \,\,=\,\,
\frac{3}{2r}\left(f(r)\right)^2,
\label{eqrgg}
\end{eqnarray}
with
\begin{eqnarray}
f(r) &=& \left\{
\begin{array}{ll}
  \arcsin^2(\sqrt{r})\,, & r \leq 1 \\
  -{1\over 4}\left[\log{1+\sqrt{1-1/r}\over 1-\sqrt{1-1/r}}-i\pi\right]^2\,,
  & r > 1 \,.
\end{array}
\right.
\end{eqnarray}

Although these functions are quite simple an analytic integration is
hard to perform. So we adopt the same procedure as for the axial-vector
case and expand $R_{gg}^{(2),\kappa}(s)$ for $s\to0$, where we take
$200$ terms into account, and for $s\to4m^2$, where $50$ terms are enough
to get a precision of $17$ to $18$ digits in the interval
$\delta\in[0.65,0.80]$.

In analogy to Eq.~(\ref{renormbeda}) the overall renormalization
condition reads:
\begin{equation}\label{renormbedk}
\lim_{q^2\to 0}\left(\Pi_S^{(2),\kappa}(q^2) -
\hat\Pi_S^{(2),\kappa}(q^2)\right) = 0\,.
\end{equation}
In this scheme, the low-energy expansion of $\Pi^{(2),\kappa}_S(q^2)$
looks as follows:
\begin{eqnarray}
C_{S,0}^{(2),s} &=& K^s \nonumber\\
C_{S,1}^{(2),s} &=&
             {4609\over 2880} 
           + {721\over 1152}\,\zeta_3
           - {4\over 9}\,\logqmms \nonumber \\
C_{S,2}^{(2),s} &=&
             {2719121\over 5806080} 
           + {10871\over 36864}\,\zeta_3
           - {28\over 135}\,\logqmms \nonumber \\ 
C_{S,3}^{(2),s} &=&
             {519513881\over 3483648000} 
           + {1330021\over 7372800}\,\zeta_3
           - {1543\over 14175}\,\logqmms \nonumber \\ 
C_{S,4}^{(2),s} &=&
             {2460910303\over 57480192000} 
           + {50939\over 409600}\,\zeta_3
           - {904\over 14175}\,\logqmms \nonumber \\ 
C_{S,5}^{(2),s} &=&
             {8958934229477\over 2929190584320000} 
           + {2526649\over 27525120}\,\zeta_3
           - {221416\over 5457375}\,\logqmms \nonumber \\ 
C_{S,6}^{(2),s} &=&
           - {37498822356303853\over 2999491158343680000} 
           + {5991294557\over 84557168640}\,\zeta_3
           - {5844896\over 212837625}\,\logqmms \nonumber \\ 
C_{S,7}^{(2),s} &=&
           - {112110439141686419569\over 6118961963021107200000}
           + {245566743541\over 4348654387200}\,\zeta_3
           - {20750416\over 1064188125}\,\logqmms\nonumber \\
C_{S,8}^{(2),s} &=&
           - {14445289941190001679673\over 723397280961606451200000}
           + {1604104532801\over 34789235097600}\,\zeta_3
           - {140071424\over 9741414375}\,\logqmms\,, \nonumber
\\&&\\
C_{S,0}^{(2),p} &=& K^p \nonumber\\
C_{S,1}^{(2),p} &=&
             {55\over 16} 
           + {175\over 96}\,\zeta_3
           - \logqmms \nonumber\\
C_{S,2}^{(2),p} &=&
             {20143\over 11520} 
           + {5047\over 4608}\,\zeta_3
           - {2\over 3}\,\logqmms \nonumber\\
C_{S,3}^{(2),p} &=&
             {2468869\over 2419200} 
           + {3969\over 5120}\,\zeta_3
           - {7\over 15}\,\logqmms \nonumber\\
C_{S,4}^{(2),p} &=&
             {920009009\over 1393459200} 
           + {974281\over 1638400}\,\zeta_3
           - {328\over 945}\,\logqmms \nonumber\\
C_{S,5}^{(2),p} &=&
             {5898858645227\over 12875563008000} 
           + {753259\over 1572864}\,\zeta_3
           - {3832\over 14175}\,\logqmms \nonumber\\
C_{S,6}^{(2),p} &=&
             {5523023003231\over 16531587072000} 
           + {11702895\over 29360128}\,\zeta_3
           - {3776\over 17325}\,\logqmms \nonumber \\  
C_{S,7}^{(2),p} &=&
             {63171728144529503\over 249957596528640000} 
           + {205215857\over 603979776}\,\zeta_3
           - {4266896\over 23648625}\,\logqmms \nonumber \\  
C_{S,8}^{(2),p} &=&
             {481264894165689829721\over 2447584785208442880000} 
           + {513043585411\over 1739461754880}\,\zeta_3
           - {32459264\over 212837625}\,\logqmms\,,
\end{eqnarray}
where $\Pi_S^{(2),\kappa}(q^2)$ and $C_{S,n}^{(2),\kappa}$ are defined
in analogy to Eqs.~(\ref{eqpis2}) and (\ref{eqpissma}), and
\begin{eqnarray}
K^s \,\,=\,\,  0.62280338337755\ldots\,\,,
&&
K^p \,\,=\,\   1.81359971877046\ldots\,\,.
\end{eqnarray}

The high energy
expansion terms up to ${\cal O}(1/z^4)$ are already listed in
\cite{HarSte97} in the $\overline{\mbox{MS}}$ scheme.  We have added the
$(m^2/s)^5$ and $(m^2/s)^6$ mass correction terms.  In the renormalization
scheme defined in Eq.~(\ref{renormbedk}) the result reads ($\logqmums =
\ln(-q^2/\mu^2)$, with $\mu$ being the renormalization scale):
\begin{eqnarray}
   \Pi^{(2),s}_{\it S} &=& {3\over 16 \pi^2} \bigg\{
K^s
-\frac{5}{8} 
-\frac{49}{16}\zeta_3
       + {m^2\over q^2}\, \bigg[
          68
          + 2\,\zeta_3
          - 20\,\zeta_5
          - 24\,\logqmums
          \bigg]
\nonumber\\&&\mbox{}
       + \left({m^2\over q^2}\right)^{2} \, \bigg[
          - 84
          + 8\,\zeta_3
          + 160\,\zeta_5
          + \left(- 36
          + 72\,\zeta_3 \right) \logqmms
          \bigg]
\nonumber\\&&\mbox{}
       + \left({m^2\over q^2}\right)^{3} \, \bigg[
          {37\over 8}
          - 62\,\zeta_3
          + 320\,\zeta_5
          + \left(- {3\over 4}
          - 36\,\zeta_3 \right) \logqmms
          + 33\,\logqmms^2
          + 12\,\logqmms^3
          \bigg]
\nonumber\\&&\mbox{}
       + \left({m^2\over q^2}\right)^{4} \, \bigg[
          {178423\over 243}
          - {4472\over 9}\,\zeta_3
          + \left({22289\over 81}
          + 16\,\zeta_3 \right) \logqmms
\nonumber\\&&\mbox{\hspace{0.4cm}}
          - 26\,\logqmms^2
          - {28\over 3}\,\logqmms^3
          \bigg]
\nonumber\\&&\mbox{}
       + \left({m^2\over q^2}\right)^{5} \, \bigg[
          {12256783\over 62208}
          - {8551\over 18}\,\zeta_3
          + \left({1594853\over 5184}
          + 46\,\zeta_3\right)\,\logqmms
\nonumber\\&&\mbox{\hspace{0.4cm}}
          + {1697\over 72}\,\logqmms^2
          + {236\over 9}\,\logqmms^3
          \bigg]
\nonumber\\&&\mbox{}
       + \left({m^2\over q^2}\right)^{6} \, \bigg[
         {474209987\over 1620000}
         - {58672\over 225}\,\zeta_3
         + \left({56656079\over 202500}
         + {576\over 5}\,\zeta_3\right)\,\logqmms
\nonumber\\&&\mbox{\hspace{0.4cm}}
         - {50407\over 90}\,\logqmms^2
         + {5758\over 27}\,\logqmms^3
         \bigg]
\label{eqpi2ss}
\bigg\}
 + \ldots\,\,,
%
\\
   \Pi^{(2),p}_{\it S} &=& {3\over 16 \pi^2} \bigg\{
   K^p
       -\frac{21}{4}\zeta_3 
       + {m^2\over q^2}\, \bigg[
          - 16\,\zeta_3
          - 20\,\zeta_5
          \bigg]
\nonumber\\&&\mbox{}
       + \left({m^2\over q^2}\right)^{2} \, \bigg[
          - 44
          + 24\,\zeta_3
          + \left( - 12
          - 72\,\zeta_3\right) \logqmms
          \bigg]
\nonumber\\&&\mbox{}
       + \left({m^2\over q^2}\right)^{3} \, \bigg[
          {221\over 8}
          + 114\,\zeta_3
          + \left( - {363\over 4}
          - 36\,\zeta_3\right) \logqmms
          - 63\,\logqmms^2
          - 12\,\logqmms^3
          \bigg]
\nonumber\\&&\mbox{}
       + \left({m^2\over q^2}\right)^{4} \, \bigg[
          {68146\over 243}
          + {1288\over 9}\,\zeta_3
          + \left( {7727\over 81}
          - 80\,\zeta_3\right) \logqmms
          - 86\,\logqmms^2
          - 44\,\logqmms^3
          \bigg]
\nonumber\\&&\mbox{}
       + \left({m^2\over q^2}\right)^{5} \, \bigg[
          {12754021\over 20736}
          + {787\over 6}\,\zeta_3
          + \left({1164055\over 1728}
          - 210\,\zeta_3\right)\,\logqmms
\nonumber\\&&\mbox{\hspace{0.4cm}}
          - {7\over 8}\,\logqmms^2
          - 172\,\logqmms^3
          \bigg]
\nonumber\\&&\mbox{}
       + \left({m^2\over q^2}\right)^{6} \, \bigg[
          {11857111\over 22500}
          - {15108\over 25}\,\zeta_3
          + \left({40729829\over 15000}
          - {3024\over 5}\,\zeta_3\right)\,\logqmms
\nonumber\\&&\mbox{\hspace{0.4cm}}
          + {11417\over 10}\,\logqmms^2
          - 738\,\logqmms^3
          \bigg]
\bigg\}
 + \ldots\,\,.
\label{eqpi2sp}
\end{eqnarray}
It should be noted that in the scalar case the singlet 
polarization function explicitly depends on the renormalization scale
$\mu$. For the approximation procedure the choice $\mu^2=m^2$
will be adopted. The imaginary part, however, is independent of $\mu$.

The analogue of Eq.~(\ref{eqpiamod}) reads
\begin{eqnarray}
  \Pi_{S,mod}^{(2),\kappa}(q^2) &=& \Pi_S^{(2),\kappa}(q^2) -
  \hat\Pi_S^{(2),\kappa}(q^2)\,.
\label{eqpikmod}
\end{eqnarray}
The leading threshold term of this function both for $\kappa = s$ and
$p$ will be determined in complete analogy to the axial-vector case (see
Eq.~(\ref{eqpismodthr})), despite the fact that for the non-singlet part
in the pseudo-scalar case it originates from the $S$-wave solution of the
Coulomb potential. If by this procedure an essential contribution to the
threshold part is missing, different Pad\'e results should develop a
large spread close to $v=0$.  One obtains
\begin{eqnarray}\label{eqpispmod1}
\Pi_{S,mod}^{(2),s}(q^2) \,\,=\,\, {3\over 16\pi^2}\ln\left({1\over
  1-z}\right)
+\ldots\,\,,
\,\,
\Pi_{S,mod}^{(2),p}(q^2) \,\,=\,\, {3\over 16\pi^2}\ln\left({1\over
  1-z}\right) \, {\pi^4\over 16}
+\ldots\,\,.
\end{eqnarray}

The approximation procedure is applied in complete
analogy to the axial-vector correlator. As for the scalar and
pseudo-scalar case also the eighth moment, $C_{S,8}^{(2),s/p}$, 
is available, Pad\'e's like $[7/7]$, $[8/6]$ or $[6/8]$ may 
be evaluated.

\section{Results}
\label{secres}

\begin{figure}[tf]
 \begin{center}
 \begin{tabular}{cc}
   \small (a) & \small (b) \\[-2ex]
   \leavevmode
   \epsfxsize=6.5cm
   \epsffile[110 270 480 560]{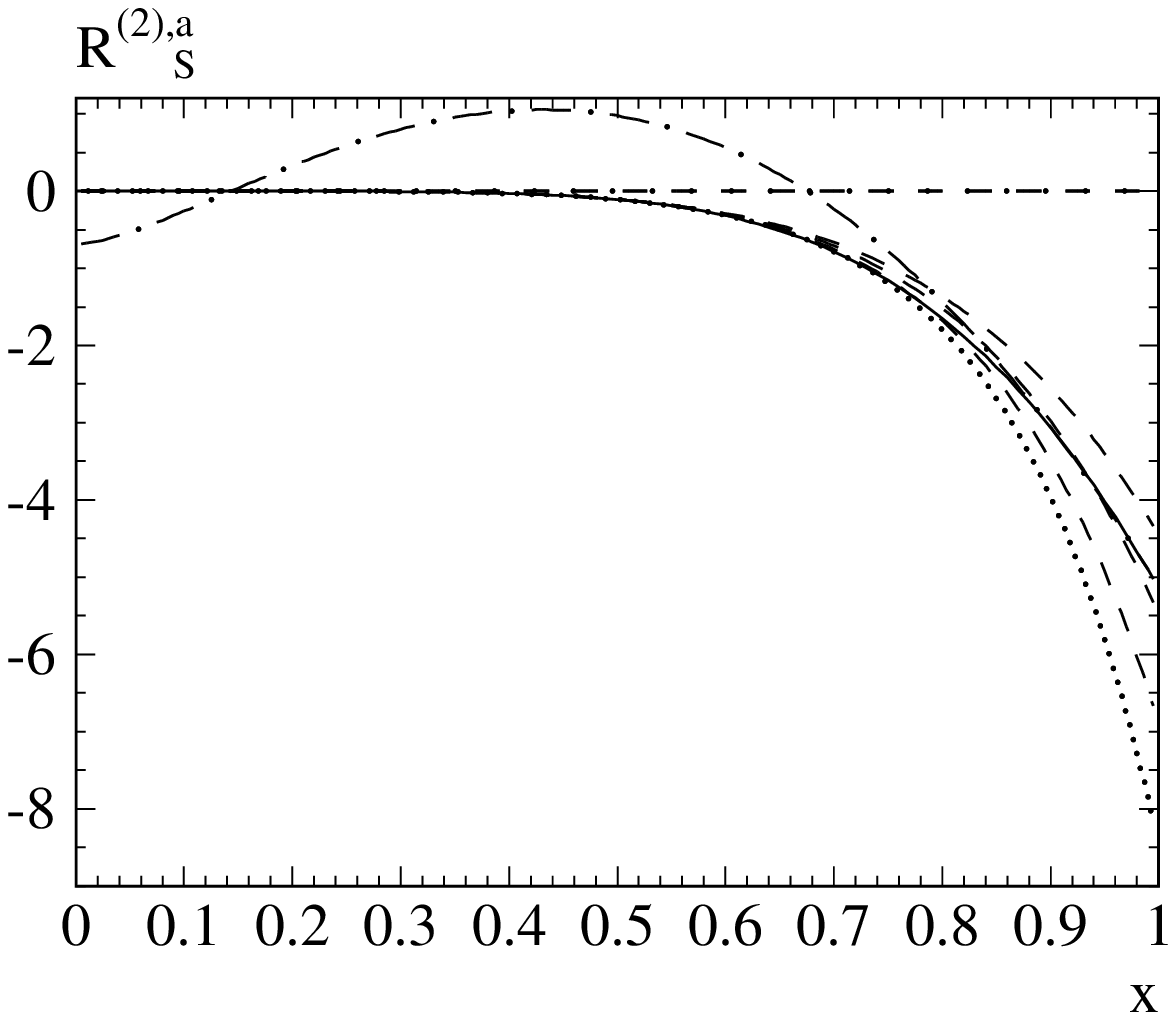 }
&
   \epsfxsize=6.5cm
   \epsffile[110 270 480 560]{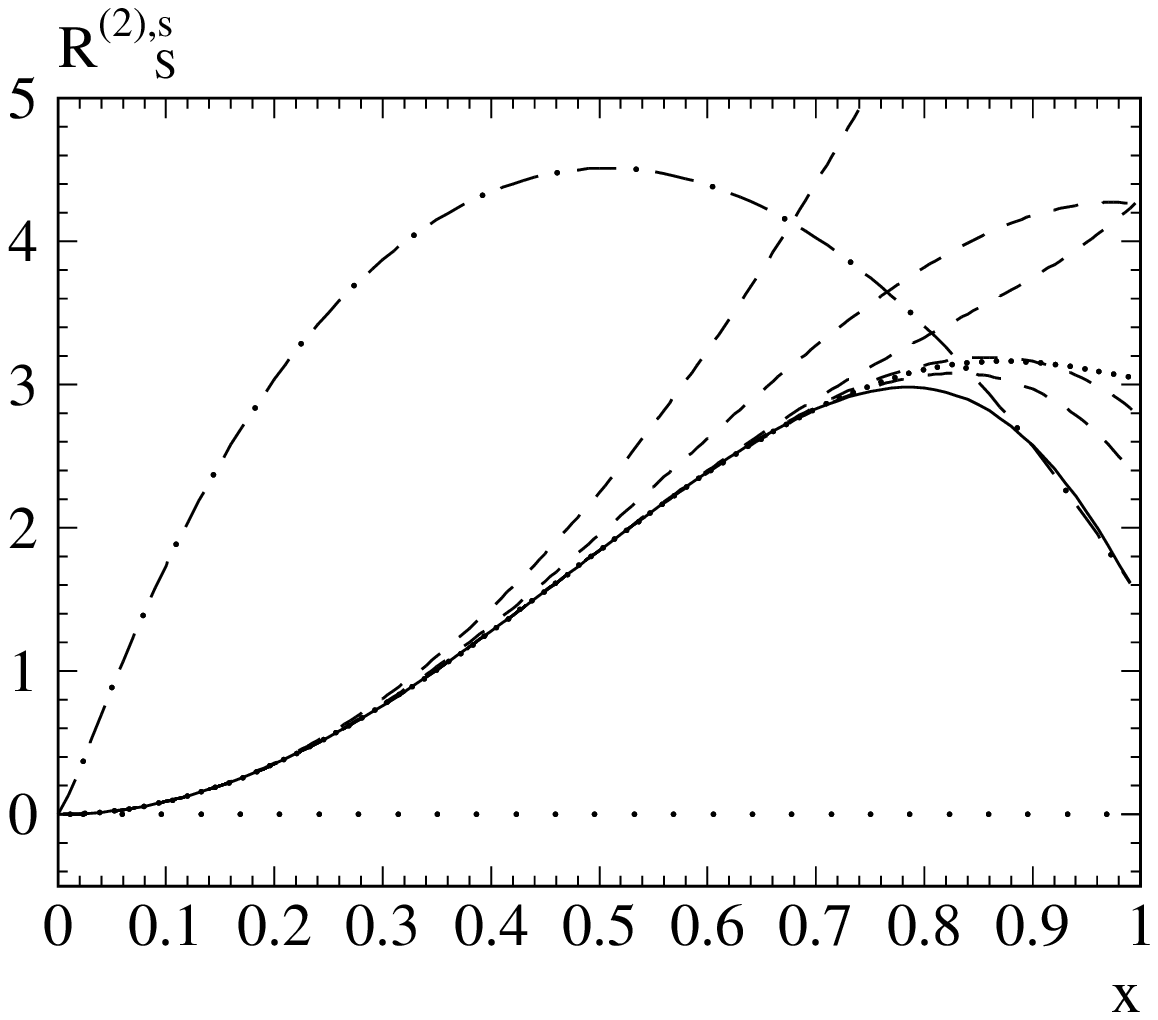 }
\\
   \small (c) & \small (d)\\[-2ex]
   \epsfxsize=6.5cm
   \epsffile[110 270 480 560]{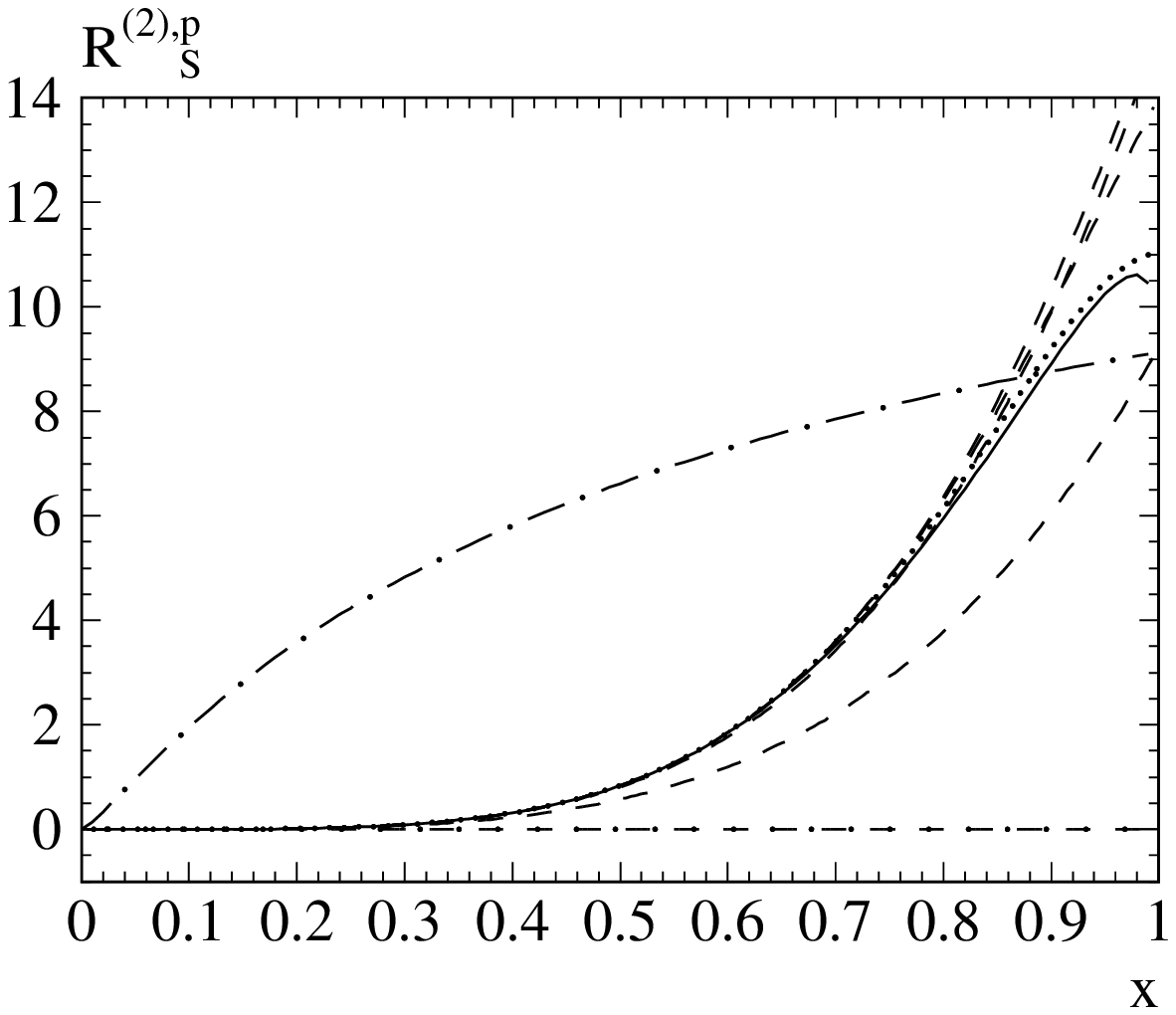 }
&
   \epsfxsize=6.5cm
   \epsffile[110 270 480 560]{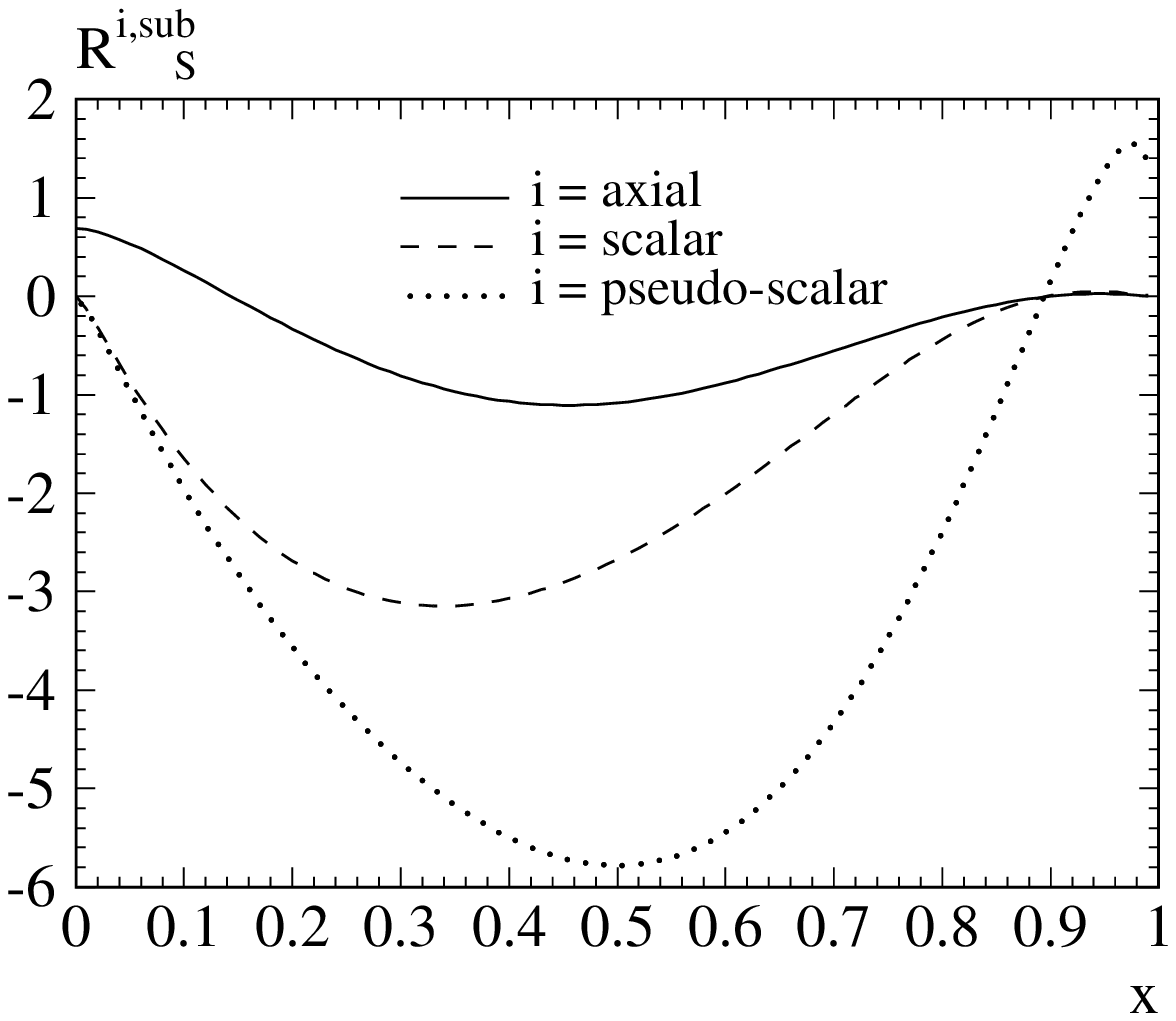 }
 \end{tabular}
 \caption[]{\label{pademxfig} The imaginary parts $R_S^{(2),a}$,
   $R_S^{(2),s}$ and $R_S^{(2),p}$ of (a) the axial-vector, (b)
   scalar and (c) pseudo-scalar singlet diagrams, respectively.
   For the abscissa the variable $x=2m/\sqrt s$ is chosen.
   Solid line: Pad\'e result; wide dots, dashes and narrow dots:
   $(m^2/q^2)^n$-expansion for $n=0$, $n=1,\ldots,5$ and $n=6$,
   respectively; dash-dotted line: purely massless cuts $R_{Sb}^{(2),a}$,
   $R_{gg}^{(2),s}$ and $R_{gg}^{(2),p}$. (d) shows the difference
   between the solid and the dashed dotted line (i.e., the contribution
   of the massive quarks) of (a), (b) and (c) as
   solid, dashed and dotted line, respectively.}
 \end{center}
\end{figure}

In Fig.~\ref{pademxfig}~(a)--(c) the results for the imaginary part of
$\Pi_{S}^{(2),a}$, $\Pi_{S}^{(2),s}$ and $\Pi_{S}^{(2),p}$ (solid
lines), together with the first seven terms of the high energy expansion
(dashed and dotted lines) are shown as functions of $x=2m/\sqrt s$.
Recall that in the displayed region, $0<x<1$, ${\rm Im}\Pi_{S}^{(2)} =
{\rm Im}\Pi_{S,mod}^{(2)}$.  Therefore, if one is interested, e.g., in
(inclusive) production of the heavy quarks only, the corresponding
massless cuts (depicted as dash-dotted lines)
have to be subtracted. 
The resulting curves are shown in Fig.~\ref{pademxfig}~(d).

As was already pointed out in \cite{HarSte972}, the imaginary part of
the axial-vector singlet contribution starts at ${\cal O}(m^6/s^3)$.
Therefore, $R_S^{(2),a}$ is rather small below $x = 1/2$.  On the other
hand, above this value the four-particle threshold at $\sqrt{s} = 4 m$
is expected to inhibit convergence of the high energy expansion.
However, there is still an agreement of the $(m^2/s)^6$-terms with the
semi-analytical Pad\'e result up to $x\approx 0.7$.  Also in the scalar
and pseudo-scalar case convergence of the high energy expansion is quite
fast up to $x\approx 0.7\ldots0.8$.  For the pseudo-scalar case even
very close to threshold the difference to the Pad\'e result is rather
small.

\begin{figure}[t]
  \begin{center}
    \begin{tabular}{cc}
   \small (a) & \small (b) \\[-2ex]
      \leavevmode
      \epsfxsize=6.5cm
      \epsffile[110 270 480 560]{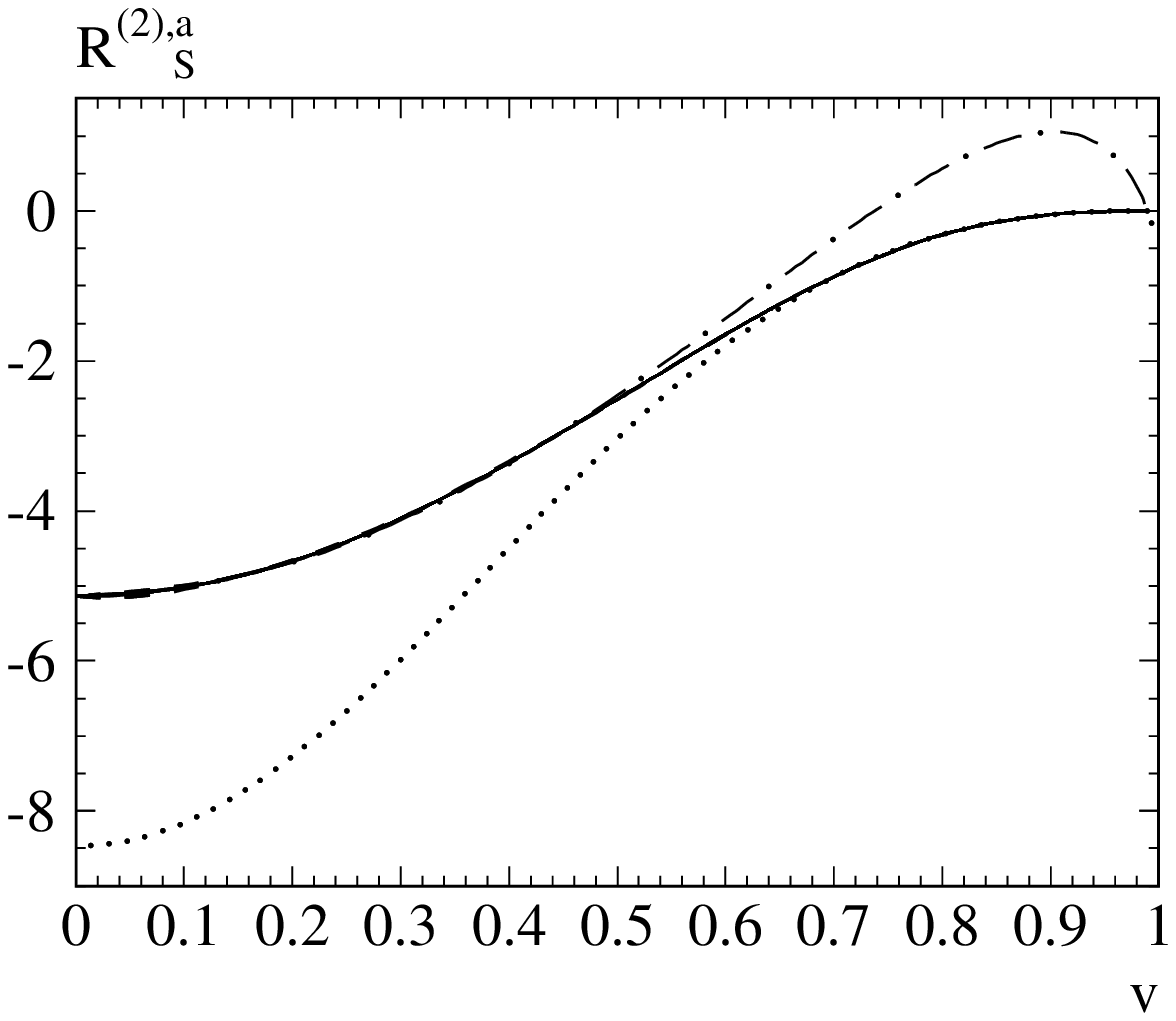 }
      &
      \epsfxsize=6.5cm
      \epsffile[110 270 480 560]{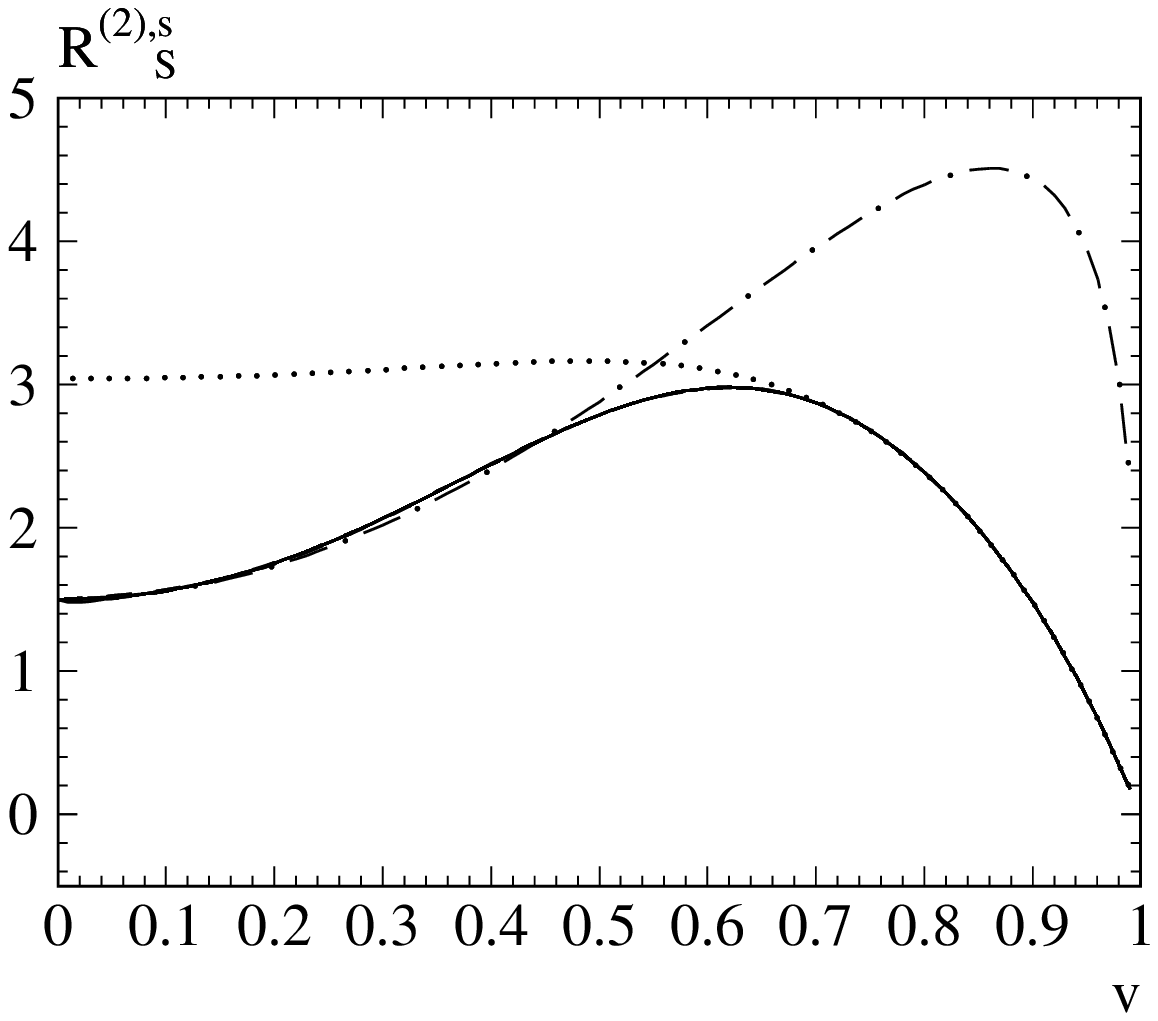 }
      \\
   \small (c) & \small (d)\\[-2ex]
      \epsfxsize=6.5cm
      \epsffile[110 270 480 560]{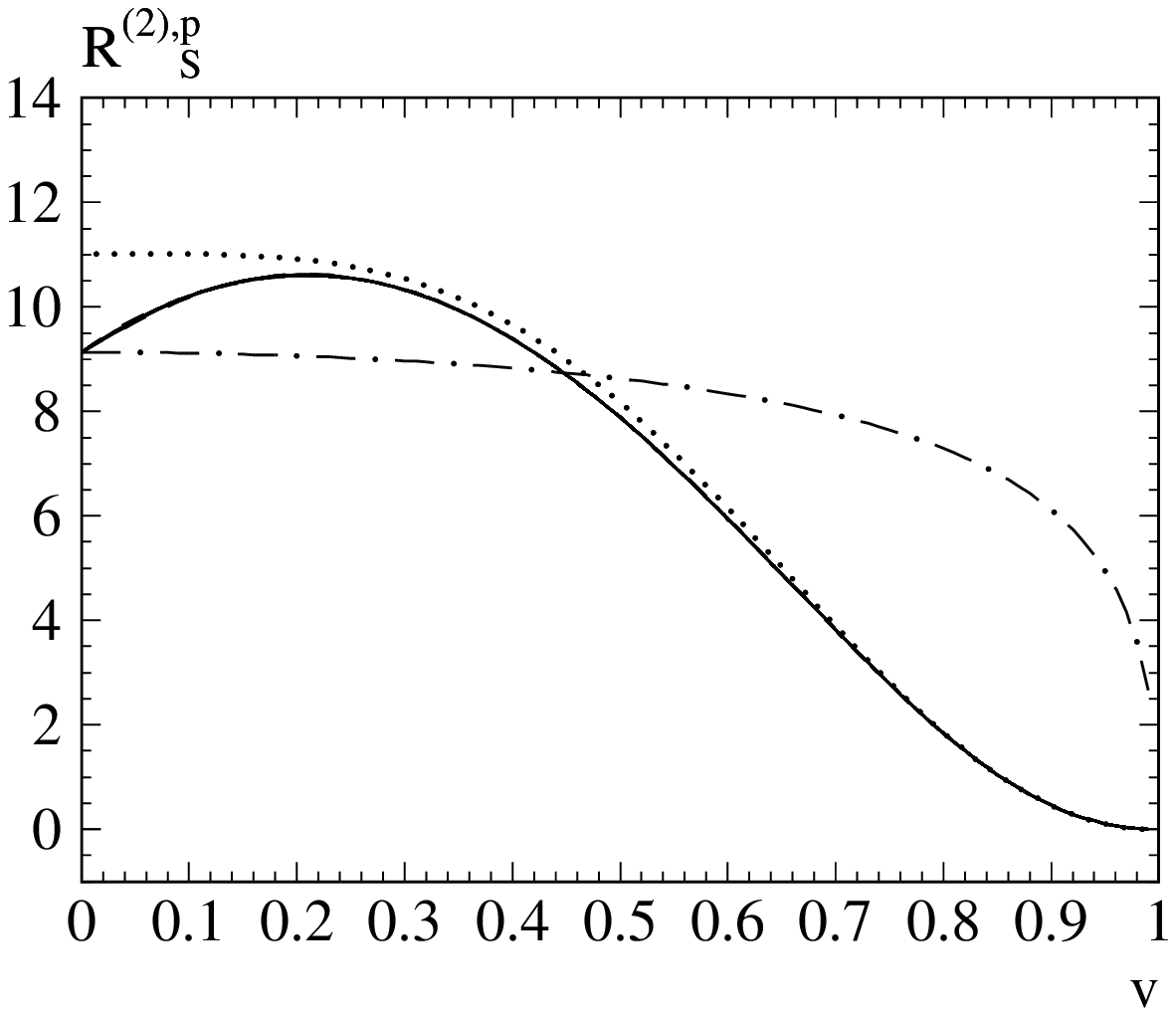 }
      &
      \epsfxsize=6.5cm
      \epsffile[110 270 480 560]{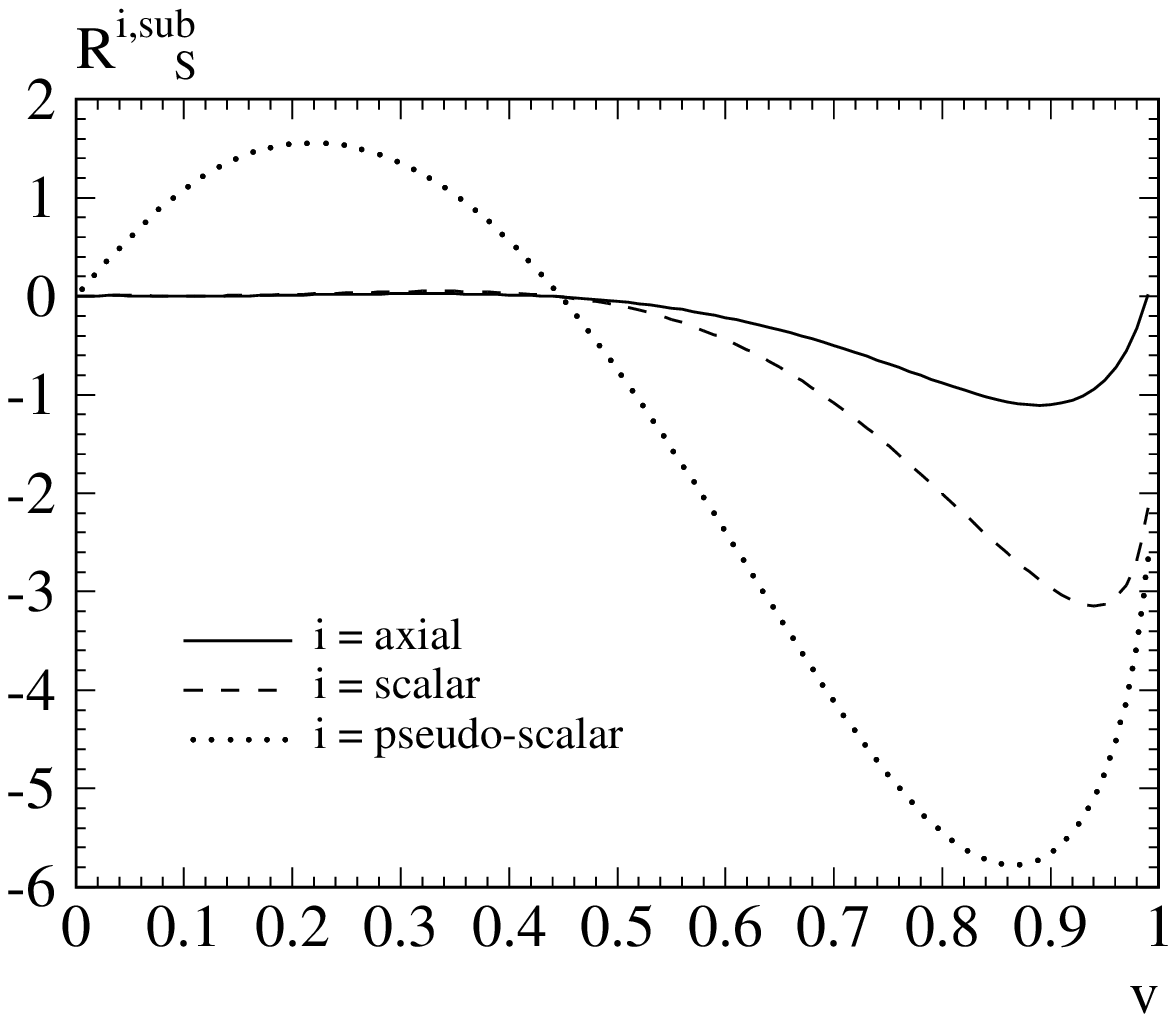 }
    \end{tabular}
    \caption[]{\label{pademvfig} $R_S^{(2),a}$, $R_S^{(2),s}$,
      $R_S^{(2),p}$ as functions of $v=\sqrt{1-4m^2/s}$.  Solid and
      dashed lines: 30 different Pad\'e curves; dotted: high energy
      expansion up to $(m^2/s)^6$; dash-dotted: purely massless cuts.
      (d) is the analogue of Fig.~\ref{pademxfig}~(d), plotted over $v$.}
  \end{center}
\end{figure}

Minor differences among the various Pad\'e approximations are visible
only close to threshold. In Fig.~\ref{pademvfig}~(a)--(c) for each
correlator 30 Pad\'e approximants with different input from $q^2\to 0$
and $q^2\to-\infty$ are plotted as functions of the quark velocity
$v=\sqrt{1-4m^2/s}$.\ The dotted curve shows the $(m^2/s)^6$-terms of
the high energy expansion for comparison.  The spread of the different
Pad\'e approximations serves as a measure for the uncertainty of the
procedure.  The dash-dotted line again corresponds to the purely
massless cuts which largely dominate the singlet contribution in all
three cases.  For the axial-vector and the scalar case the differences
$R_S^{(2),a}-R_{Sb}^{(2),a}$, respectively, $R_S^{(2),s}-R_{gg}^{(2),s}$
are remarkably small below $v=0.5$ which confirms our previous
assumptions on the threshold behavior in these cases
(see Fig.~\ref{pademvfig}~(d)). As was expected the analogue quantity in the
pseudo-scalar case grows steeper for $v\to0$, but still the stability of
the Pad\'e result justifies the threshold assumption.  We even modified
artificially Eq.~(\ref{eqpispmod1}) by $\pm10$\% and observed a drastic
destabilisation of the Pad\'e approximants.  This also gives quite some
confidence to the ansatz for $\Pi_S^{(2),p}$ at threshold.

In the following handy approximation formulae for the imaginary parts
$R_S^{(2),a}, R_S^{(2),s}$ and $R_S^{(2),p}$ are given. To keep the
formulae as short as possible a Pad\'e approximant containing the high
energy input up to ${\cal O}(m^2/s)$ is selected. After subtraction of
the known analytic results the (small) remainder, which vanishes both
for $v=0$ and $v=1$ is approximated using Legendre polynomials.  The
results read:
\begin{eqnarray}
R_S^{(2),a} &=&
-\frac{57}{8} + \frac{9}{2}\ln4 - \frac{9}{2}\zeta_2 + \frac{21}{8}\zeta_3
+v\left(\frac{171}{16}-\frac{27}{4}\ln4\right)
+v^3\left(-\frac{57}{16}+\frac{9}{4}\ln4\right)
\nonumber\\&&\mbox{}
+\left(\frac{27}{4}v-\frac{9}{4}v^3\right)\zeta_2
+\left(-\frac{63}{16}v + \frac{21}{16}v^3\right)\zeta_3
\nonumber\\&&\mbox{}
+ 12\left(\frac{m^2}{s}\right)^{3/2}\left(v^4 - 2 v^2\right)^4
\nonumber\\&&\mbox{}
+ 50p^{3/2} (1-p) \left[
-0.0789\, P_0(p)
-0.0869\, P_1(p)
+0.1099\, P_2(p)
\right],
\\
R_S^{(2),s} &=&
\frac{3}{2}
+\frac{27}{4}v-\frac{33}{4}v^3
\nonumber\\&&\mbox{}
+ 3\left(\frac{m^2}{s}\right)^{3/2}\left(v^4 - 2 v^2\right)^4
\nonumber\\&&\mbox{}
+ 50p^{3/2} (1-p) \left[
-0.193\, P_0(p)
+0.111\, P_1(p)
+0.0277\, P_2(p)
\right],
\\
R_S^{(2),p} &=&
\pi^4
\left(\frac{3}{32} - \frac{9}{64}v +\frac{3}{64}v^3\right)
\nonumber\\&&\mbox{}
- 35 \left(\frac{m^2}{s}\right)^{3/2}\left(v^4-2 v^2\right)^4
\nonumber\\&&\mbox{}
+ 50p^{3/2} (1-p) \left[
1.586\, P_0(p)
-1.822\, P_1(p)
+0.525\, P_2(p)
\right]
\,,
\end{eqnarray}
with $p = (1-v)/(1+v)$.  Note again that the corresponding massless cuts
(see Eqs.~(\ref{eqrsb}) and (\ref{eqrgg})) have to be
subtracted in order to find the rates for, e.g., $t\bar t$ production.

To summarize, the polarization functions for axial-vector, scalar
and pseudo-scalar singlet current correlators have been computed. To this
end the knowledge of $\Pi(q^2)$ for $q^2\to0$, $q^2\to-\infty$
and $q^2\to4m^2$ has been exploited in combination with conformal mapping
and Pad\'e approximation. Results have been presented for the imaginary
parts and handy approximation formulae have been provided. This completes
the evaluation of three-loop polarization functions at ${\cal O}(\alpha_s^2)$
as in \cite{CheKueStePade} the non-singlet contributions were considered.

\vspace{2mm} \centerline{\bf Acknowledgments} 
\medskip\noindent
We would like to thank B.K. Kniehl and J.H. K\"uhn 
for helpful discussions. MS appreciates the warm
hospitality at the Institute for Theoretical Particle Physics at the
University of Karlsruhe.  This work was supported by BMBF under Contract
057KA92P, DFG under Contract Ku 502/8-1 and INTAS under Contract
INTAS-93-744-ext. RH thanks the ``Landesgraduiertenf\"orderung'' at the
University of Karlsruhe for financial support.


\end{document}